# Monolithic Integrated Multiband Acoustic Devices on Heterogeneous Substrate for Sub-6 GHz RF-FEMs

Shibin Zhang[1,3,#,*], Hongyan Zhou[1,2,#], Pengcheng Zheng[1,2], Jinbo Wu[1,2], Liping Zhang[1,2], Zhongxu Li[1,2], Kai Huang[1], Xin Ou[1,2,*], Xi Wang[1,2]

[1]State Key Laboratory of Functional Materials for Informatics, Shanghai Institute of Microsystem and Information Technology, Chinese Academy of Sciences, Shanghai 200050, China

[2]Center of Materials Science and Optoelectronics Engineering, University of Chinese Academy of Sciences, Beijing 100049, China

[3]Xin Ou Integration Technology Co., Ltd (XOITEC), Shanghai 201899, China

*Corresponding authors: sbzhang@mail.sim.ac.cn; ouxin@mail.sim.ac.cn; #Authors contributed equally to this work.

(October 20, 2021)

*Abstract*— Monolithic integration of multiband (1.4~ 6.0 GHz) RF acoustic devices were successfully demonstrated within the same process flow by using the lithium niobate (LN) thin film on silicon carbide (LNOSiC) substrate. A novel surface mode with sinking energy distribution was proposed, exhibiting reduced propagation loss. Surface wave and Lamb wave resonators with suppressed transverse modes and leaky modes were demonstrated, showing scalable resonances from 1.4 to 5.7 GHz, electromechanical coupling coefficients ($k^2$) between 7.9% and 29.3%, and maximum Bode-$Q$ ($Q_{max}$) larger than 3200. Arrayed filters with a small footprint (4.0×2.5 mm$^2$) but diverse center frequencies ($f_c$) and 3-dB fractional bandwidths (FBW) were achieved, showing $f_c$ from 1.4 to 6.0 GHz, FBW between 3.3% and 13.3%, and insertion loss (IL) between 0.59 and 2.10 dB. These results may promote the progress of hundred-filter sub-6 GHz RF front-end modules (RF-FEMs).

## I. Introduction

Radio-frequency acoustic filters are essential components of the front-ends for emerging applications in 5G [1-3]. The number of filters in the RF-FEM chipset of a high-end smartphone continues to grow, and may exceed 100 soon [4, 5]. Therefore, high-density RF-FEMs with greater number of filters and smaller footprints are in critical demand to meet the performance requirements and to fit the miniaturization of smartphone area. Fig. 1 shows the schematic of a RF-FEM, in which die-level surface acoustic wave (SAW) and bulk acoustic wave (BAW) filters are usually integrated using highly customized flip-chip processes [4]. The packaged filters are schematically shown in Fig. 2, where the SAW devices (e.g., SAW [6, 7], TC-SAW [8, 9], IHP-SAW [10, 11], Ultra-SAW [12]) are generally applied to bands below 3 GHz while the BAW devices (e.g., FBAR [13], SMR-BAW [14], enhanced FBAR [15]) can achieve operating frequencies ($f$) higher than 3 GHz. Despite the successful commercial applications in RF-FEMs at this stage, the flip-chip packaged discrete SAW and BAW filters with different structures (on various piezoelectric substrates), designs, and incompatible manufacturing processes are costly, complicated, and large-sized.

This work explores the possibility of monolithic integration of 4G-LTE, 5G-FR1 and WiFi-6 filters with similar structures by the same process flow. Based on LNOSiC, the newly proposed surface mode devices with sinking energy distribution and the Lamb mode devices with special designs were monolithically integrated, as schematically shown in Fig. 3. In-band spurious responses of both modes were effectively mitigated. Resonators with scalable resonances (1.4~5.7 GHz) and the $Q_{max}$ larger than 3200 were achieved, showing $k^2$ between 7.9% and 29.3%. Arrayed filters with a compact footprint of 4.0×2.5 mm$^2$ show $f_c$ from 1.4 to 6.0 GHz, FBW between 3.3% and 13.3%, and IL between 0.59 and 2.10 dB.

## II. Material Platform for Multiband Devices

Since the $f$ of a BAW device is inversely proportional to the thickness of the sandwiched piezoelectric layer (e.g., AlN [16], AlScN [17,18]), BAW devices for multiple bands require the precisely controlled piezoelectric layers of various thicknesses as well as the complicated fabrication processes. Due to the limited phase velocities ($V_p$) of surface waves (e.g., ~4000 m/s of SH0 mode), high-performance SAW devices on piezoelectric heterogeneous substrates (e.g., LiNbO$_3$-on-Si, LNOSi) can hardly achieve a $f$ higher than 3.0 GHz [19, 20, 21], so as to be applied to bands above 3 GHz. Lamb wave (e.g., S0 mode) devices with the $V_p$ greater than 6000 m/s can easily achieve a $f$ higher than 3 GHz [22, 23]. Besides, S0 mode and SH0 mode devices generally have similar electrodes structure (interdigital transducers, IDT) and the same but simple fabrication process flow.

Potential material platforms for monolithic integration of SH0 mode and S0 mode devices for sub-6 GHz RF-FEMs have been comparatively studied through finite element analysis. Fig. 4 shows the simulated displacement mode shapes of the S0 mode, where the resonant energy of the S0 mode can be well confined in the LNOSiC surface while almost totally leaks into the underneath silicon of LNOSi. Fig. 5 shows the simulated temperature increment (ΔT) of different substrates when the same boundary heat sources of $10^5$ W/m$^2$ were applied on their top surfaces. By comparison, integrating LiNbO$_3$ thin films onto substrates with higher thermal conductivity can decrease the ΔT effectively, while the SiO$_2$ buried layer will increase the interface thermal resistance, resulting in a higher ΔT. Therefore, LNOSiC is an excellent substrate possessing the advantages of both acoustic energy confinement and heat dissipation.

### III. DEVICE DESIGN AND FABRICATION

#### A. Dispersion of S0 mode and SH0 mode

Based on LNOSiC, dispersion curves of the $f$, $V_p$, and $k^2$ of the SH0 and S0 modes were simulated and shown in Fig. 6, in which $\lambda$ is the wavelength of the intended mode. The thickness of the LiNbO$_3$ thin film ($h$) and the Al electrode are set to be 450 nm and 120 nm, respectively. Both modes show large $k^2$, which is proportional to filter bandwidth. The $V_p$ of the S0 mode is about 1.5 times greater than that of the SH0 mode, so as the $f$. Therefore, the S0 mode devices will be preferentially applied to bands above 3 GHz, while the SH0 mode devices to bands below 3 GHz. As an example, Figs. 7(a)-(f) show the simulated admittance curves and the corresponding displacement mode shapes of one-port resonators with resonance frequencies ($f_r$) lower than 2 GHz, 3 GHz, and higher than 3 GHz.

#### B. SH0 mode with sinking energy distribution (SH0-SED)

Due to the dispersion of acoustic modes, a heterogeneous substrate with thickness matched layers usually corresponds to SAW devices (e.g., IHP-SAW) of a specific band. Thanks to the extremely high velocity of the slow shear bulk wave ($V_{SSB}$, ~7150 m/s) and the excellent $f\cdot Q$ product of SiC, the SH0 mode devices on LNOSiC with scalable resonances from 1.3 to 3.0 GHz ($h/\lambda<0.25$) all maintain large admittance ratios (see examples in Figs. 7(a) and (b)). Especially, when $h/\lambda$ is less than 0.15, the SH0 mode with sinking energy distribution (SH0-SED) exhibits even lower propagating loss. As an example, Fig. 8(a) shows the schematic of the delay line structure used for comparing the propagating loss of the SH0-SED ($h/\lambda=0.09$) and the SH0 ($h/\lambda=0.18$) modes, where the damping loss (Λ) of the LiNbO$_3$ is set to be 0.002. Figs. 8(b) and (c) show the simulated S21 parameters, from which the loss factors (δ) were extracted and plotted in the insets. The smaller $h/\lambda$ gets the lower δ, which will contribute to high-$Q$ SH0-SED mode devices.

#### C. Suppression of the in-band spurious modes

As shown in Figs. 7(a)-(c), both modes are suffered from the transverse mode spurious responses (transverse standing waves) while the S0 mode is also affected by the leaky mode and the overtone modes. Fig. 9(a) shows the simulated mode shapes of the main mode and the first three-order transverse modes of a SH0 mode resonator, and Fig. 9(b) shows the corresponding admittance curve and conductance curve. By adding small border regions (piston IDT) at the tip and the root of the IDT fingers, the transverse modes can be effectively mitigated, as shown in Fig. 9(c).

Fig. 10(a) shows the simulated mode shapes of the leaky mode near the anti-resonance ($f_a$) of the S0 mode. In our opinion, the leaky mode is mainly caused by the reflectors with insufficient reflection. By appropriately reducing the pitch of the reflectors, the leaky mode can be well suppressed, as shown in Figs. 10(b)-(c). Based on the above optimized resonators, ladder-type filters with mitigated in-band spurious responses were designed, and the simulated S21 parameters were plotted in Fig. 11, showing $f_c$ from 1.4 to 4.4 GHz and FBW between 4.2% and 15.5%.

#### D. Fabrication of the LNOSiC substrate and the devices

Fig. 12(a) presents the image of the 4-inch LNOSiC wafer, which was fabricated by the ion-cutting process. Fig. 12(b) shows the cross-sectional scanning electron microscope (SEM) image of the LNOSiC, in which the thickness of the LiNbO$_3$ layer is about 470 nm. The resonators and filters were prepared through electron beam lithography, metal (Ti/Al) evaporation and lift-off process. Fig. 13 shows the optical microscope image of the fabricated filters with a footprint of 4.0×2.5 mm$^2$, and the insets show the zoom-in resonators as well as the optimized IDTs and reflectors.

### IV. MEASUREMENT RESULTS

The frequency responses of the demonstrated devices were characterized using a VNA (E5071C) at room temperature in air. Figs. 14 and 15 show the measured admittance and Bode-$Q$ curves of the low-band (SH0-SED mode) and the mid-band (SH0 mode) resonators with and without the piston IDTs, in which the transverse modes were effectively mitigated while the $Q_{max}$ is almost unaffected. As predicted in Fig. 8, the SH0-SED mode resonator exhibits an excellent $Q_{max}$ of 3680 at 1.45 GHz, resulting in a high figure-of-merit ($k^2 \cdot Q_{max}$) of above 319. Figs. 16(a) and (b) show the measured admittance curves of the high-band (S0 mode) resonators with $f_r$ of 3.7 GHz and 4.4 GHz, respectively. When the pitch of the reflector ($P_r$) is 0.9 times the pitch of the IDTs ($P_i$), the leaky modes of both S0 mode resonators were effectively suppressed. By comparison, it is found that the experimental results shown in Figs. 14-16 and the simulations in Figs. 9 and 10 are in great agreement. Figs. 17(a) and (b) show the admittance curves and the extracted $k^2$ of the prepared SH0 mode and S0 mode resonators, respectively, showing broadband resonances (1.4~5.7 GHz) and large $k^2$.

Fig. 18 shows the measured S21 parameters of the prepared monolithically integrated ladder-type filters (labeled as F1~F8), and the table below lists the corresponding IL and the 3-dB bandwidth, showing $f_c$ from 1.4 to 6.0 GHz, the minimum IL of 0.59 dB and the maximum bandwidth of 488 MHz. With optimized designs, the in-band spurious responses of the arrayed filters have been effectively mitigated. The insets in Fig. 18 serve as examples, where the zoomed-in in-band responses of the filters with and without the optimized designs are compared. The filters with SH0 mode resonators are preferred to 4G-LTE bands, while those with S0 mode resonators are suitable for 5G-FR1 and WiFi-6 bands. Note that, the spurious passband of the S0 mode filter (marked by a black dashed box) is mainly caused by the overtone modes (labeled in Fig. 7(c)) and can be mitigated by using the embedded IDTs, as the simulations shown in Fig. 19(a). The SEM image of the preliminarily prepared embedded IDTs is shown in Fig. 19(b), in which the electrode deposition process needs to be optimized.

## V. Conclusions

The monolithic integration of surface wave and Lamb wave filters for 4G-LTE, 5G-FR1 and WiFi-6 bands were firstly demonstrated on a single LNOSiC substrate. In-band spurious responses of the filters were effectively mitigated. The prepared resonators show scalable resonances from 1.4 to 5.7 GHz, and $k^2$ between 7.9% and 29.3%. The newly proposed SH0-SED mode exhibits an excellent $Q_{max}$ of 3680. Arrayed filters with a footprint of 4.0×2.5 mm$^2$ show $f_c$ from 1.4 to 6.0 GHz, FBW between 3.3% and 13.3%, and IL between 0.59 and 2.10 dB. Further optimization of this technology may promote the progress of hundred-filter sub-6 GHz RF-FEMs.

**Acknowledgment.** The authors acknowledge the support from the National Key R&D Program of China (2020YFB2008802, 2019YFB1803903), National Natural Science Foundation of China (61874128, 61851406).

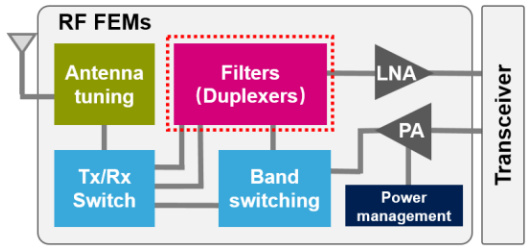
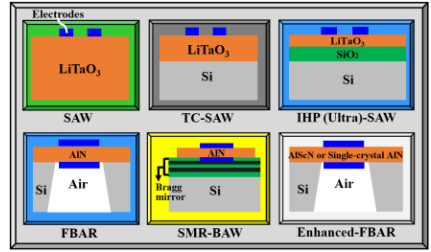
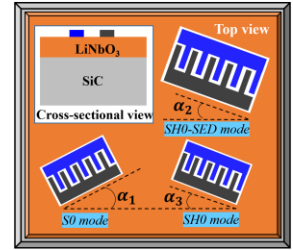

Fig. 1. The schematic of a RF front-end module (RF-FEM), in which the filters, duplexers and multiplexers (composed of filters) are key components.

Fig. 2. The schematic of the packaged die-level SAW and BAW filters using flip-chip process.

Fig. 3. The schematic of the monolithic integration of multiband filters using SAW and Lamb wave resonators on LNOSiC substrate.

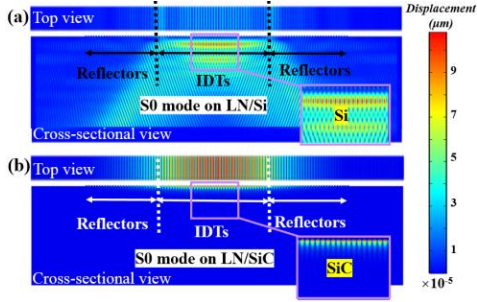
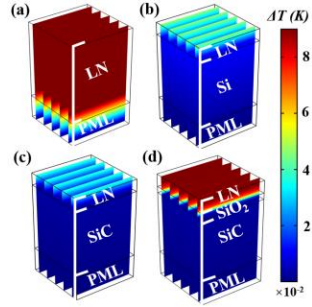
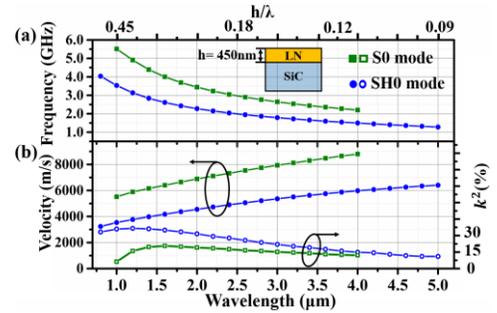

Fig. 4. The top view and the cross-sectional view of the displacement mode shapes at the resonance frequency of a one-port S0 mode resonator based on the (a) LNOSi and the (b) LNOSiC substrates.

Fig. 5. The simulated temperature increment ($\Delta T$) of the (a) LiNbO$_3$ crystal, (b) LNOSi, (c) LNOSiC and the (d) LN-SiO$_2$-SiC substrates with the top-surface boundary heat sources of $10^5$ W/m$^2$.

Fig. 6. The simulated (a) resonance frequency and (b) phase velocity ($V_p$) and $k^2$ curves of the SH0 and S0 modes based on the LNOSiC substrate as the function wavelength ($\lambda$).

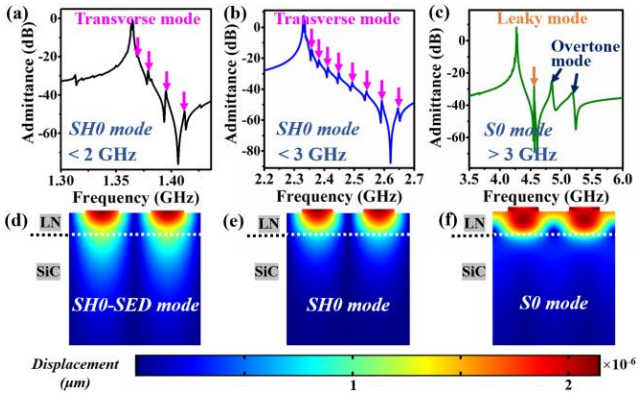
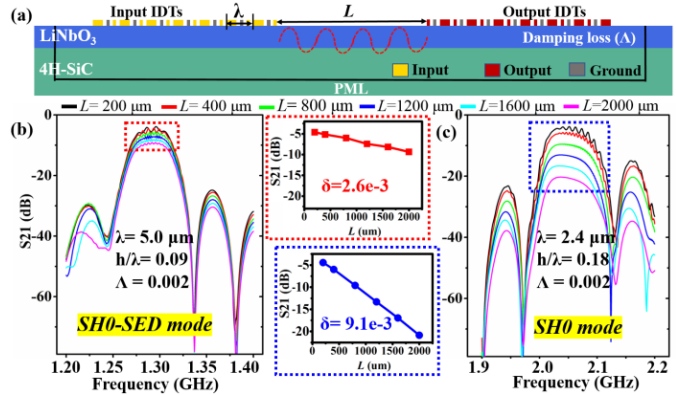

Fig. 7. (a)-(c) The simulated admittance curves and (d)-(f) the corresponding displacement mode shapes of one-port resonators with resonance frequencies lower than 2 GHz, 3 GHz, and higher than 3 GHz.

Fig. 8. (a) The schematic of the delay line structure with the key parameters listed. (b)-(c) The simulated S21 parameters of the SH0-SED and SH0 modes. The extracted loss factors ($\delta$) are plotted in the insets.

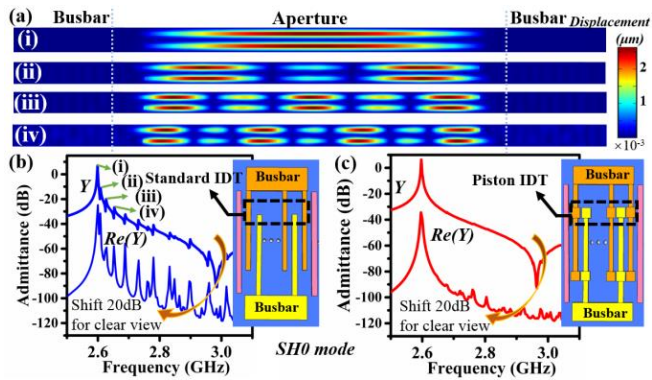
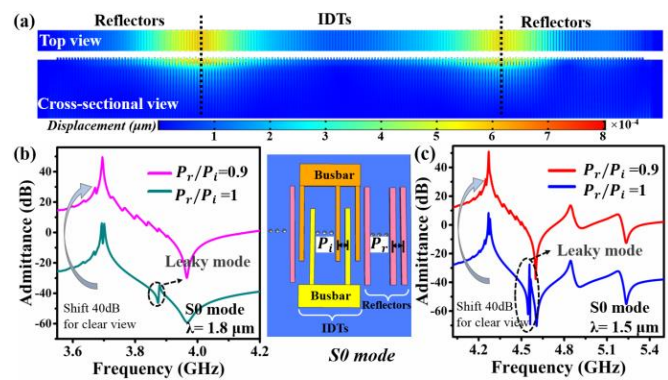

Fig. 9. (a) the simulated mode shapes of the main mode and the first three-order transverse modes of the SH0 mode resonator. The simulated admittance and conductance curves of the SH0 mode resonator with the (b) standard IDTs and the (c) piston IDTs.

Fig. 10. (a) The simulated mode shapes of the leaky mode of the S0 mode resonator. The simulated admittance responses of the S0 mode resonator with the pitch of the reflector ($P_r$) equals to be 0.9 times the pitch of the IDTs ($P_i$) when (b) $\lambda$=1.8 μm and (c) $\lambda$=1.5 μm.

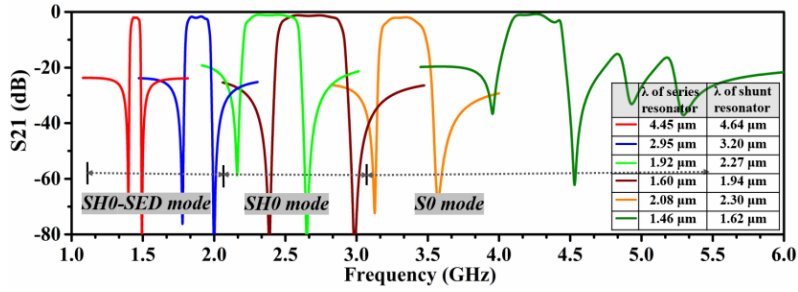
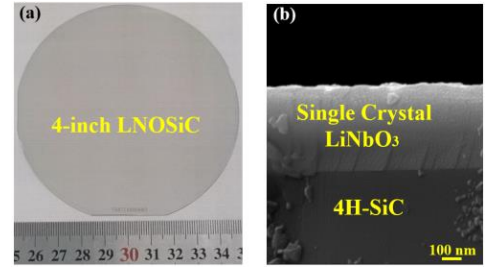

Fig. 11. The simulated S21 parameters of the multiband ladder-type filters composed of the SH0 mode and S0 mode resonators with mitigated in-band spurious responses.

Fig. 12. (a) The photo and (b) the cross-sectional SEM image of the fabricated 4-inch LNOSiC substrate.

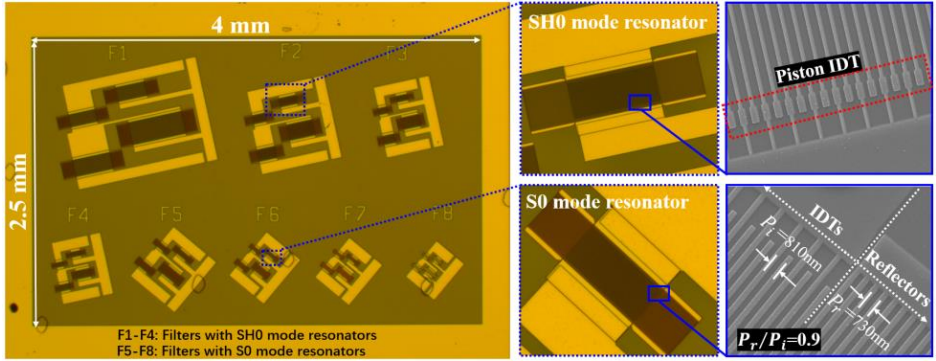
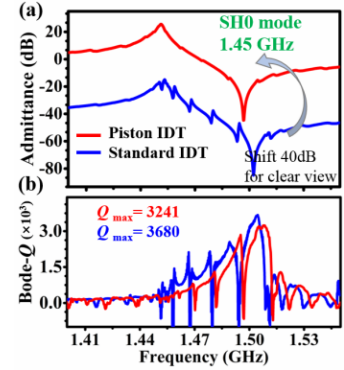

Fig. 13. The optical and the SEM images of the fabricated multiband ladder-type filters composed of the optimized SH0 mode and the S0 mode resonators.

Fig. 14. The measured (a) admittance and (b) the Bode-$Q$ curves of the fabricated SH0 mode resonators with the standard and the piston IDTs.

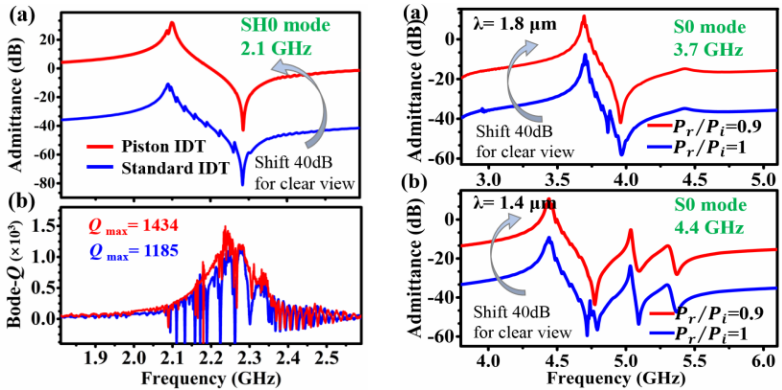
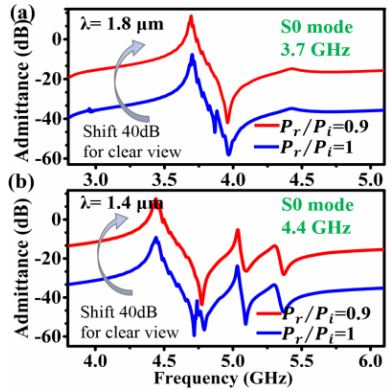
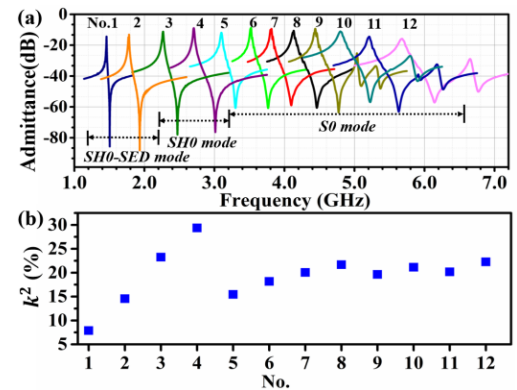

Fig. 15. The measured (a) admittance and (b) the Bode-$Q$ curves of the fabricated SH0 mode resonators with the standard and the piston IDTs.

Fig. 16. (a)-(b) The measured admittance curves of the fabricated S0 mode resonators with the standard and the shortened pitches of reflectors.

Fig. 17. (a) The measured admittance curves and the (b) extracted $k^2$ of the fabricated SH0 mode and S0 mode resonators.

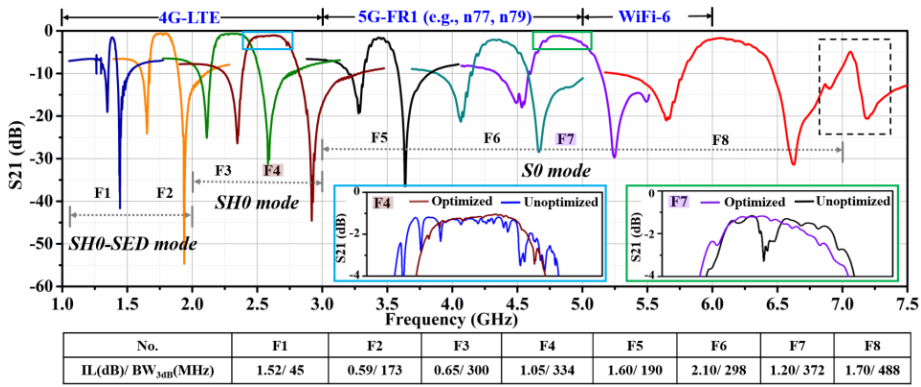
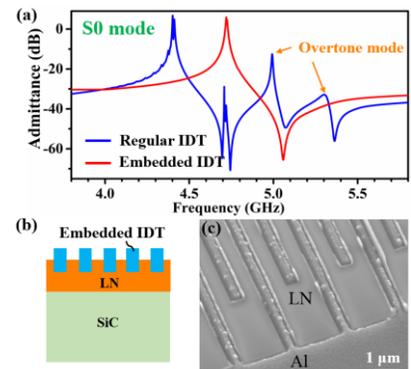

Fig. 18. The measured S21 parameters of the prepared monolithically integrated ladder-type filters (labeled as F1~F8). The insets show the zoomed-in in-band responses of the filters with and without the optimized designs while the table below lists the corresponding IL and the 3-dB bandwidth ($BW_{3dB}$).

Fig. 19. (a) The simulated admittance curves of the S0 mode resonators with and without the embedded IDTs. (b) The schematic of the embedded IDTs. (c) The SEM image of the preliminarily prepared embedded IDTs.